\documentclass{aip-cp}

\usepackage[numbers]{natbib}
\usepackage{rotating}
\usepackage{graphicx}

\begin{document}

\title{Collective Neutrino Oscillations and Nucleosynthesis}

\author[aff1]{A.B. Balantekin\corref{cor1}}
\eaddress[url]{http://nucth.physics.wisc.edu}

\affil[aff1]{University of Wisconsin, Department of Physics, Madison, Wisconsin 53706 USA}
\corresp[cor1]{baha@physics.wisc.edu}

\maketitle

\begin{abstract}
The emergent phenomenon of collective neutrino oscillations arises from neutrino-neutrino interactions in environments with very large number of neutrinos. Since such 
environments are likely sites of the heavy-element synthesis, understanding all aspects of collective neutrino oscillations seems to be necessary for a complete accounting of 
nucleosynthesis. I briefly summarize some of the salient features along with recent work on the properties and astrophysical applications of the collective neutrino oscillations. 
\end{abstract}

\vskip 1.5cm
\section{INTRODUCTION} 
All astrophysical environments where synthesis of heavier elements is supposed to occur, may they be core-collapse supernovae, merging neutron stars, or gamma-ray burst emitting objects, contain numerous neutrinos. For example a proto-neutron star, formed following the emission of a core-collapse supernova, cools down by emitting neutrinos with the total energy of $\sim 10^{59}$ MeV. If the average energy of each neutrino is $\sim 10$ MeV, this corresponds to $10^{58}$ neutrinos streaming out of the core in a short time. In such environments a proper description of the neutrino transport must include neutrino-neutrino interactions, a Standard Model process usually omitted in all other applications of weak interactions.  When the neutrino transport is dominated by the coherent forward scattering of neutrinos, the effect is proportional to $G_F$ (as opposed to $G_F^2$ for collisions when it can be safely ignored). In the case of coherent forward scattering the Hamiltonian which describes neutrino transport is given by 
\begin{equation}
\label{total}
\hat{H} 
= \left(
\sum_p\frac{\delta m^2}{2p}\mathbf{B}\cdot\mathbf{J}_p  - \sqrt{2} G_F 
N_e  J_p^0  \right) 
+ \frac{\sqrt{2}G_{F}}{V}\sum_{\mathbf{p} \neq\mathbf{q}}\left(1- 
\cos\vartheta_{\mathbf{p}\mathbf{q}}\right)\mathbf{J}_{\mathbf{p}}\cdot\mathbf{J}_{\mathbf{q}}  
\end{equation} 
where an auxiliary vector quantity 
\begin{equation}
\mathbf{B} = (\sin2\theta,0,-\cos2\theta)  
\end{equation}
is parameterized in terms of the neutrino mixing angle $\theta$. In writing these equations, for simplicity we ignored antineutrinos and assumed  only two flavors. 
It is straightforward to relax this assumption  \cite{Balantekin:2006tg}. 
Eq. (\ref{total}) is written using the neutrino flavor isospin algebras: 
\begin{eqnarray}
J^+_{{\bf p}} &=& a_x^\dagger({\bf p}) a_e({\bf p}), \> \> \>
J^{\>-}_{{\bf p}}=a_e^\dagger({\bf p}) a_x({\bf p}), \nonumber \\
J^0_{{\bf p}} &=& \frac{1}{2}\left(a_x^\dagger({\bf p})a_x({\bf p})-a_e^\dagger({\bf p})a_e({\bf p}) 
\right). \label{su2}
\end{eqnarray}
where neutrino creation and annihilation operators, $a_e, a_e^\dagger, a_x, a_x^\dagger$ are introduced.  These operators span SU(2)  algebras labeled by each neutrino momenta.
In Eq. (\ref{total}) $\cos\vartheta_{\mathbf{p}\mathbf{q}}$ is the angle between neutrino momenta $\mathbf{p}$ and $\mathbf{q}$ and $N_e$ is the electron density in the medium. 
Unlike the one-body Hamiltonian of the
matter-enhanced neutrino oscillations where neutrinos interact with a mean-field
(generated by the background particles other than neutrinos), the Hamiltonian
describing the many-neutrino gas in a core-collapse supernova contains both one- and
two-body terms, making it technically much more challenging. Inclusion of the
neutrino-neutrino interaction terms leads to very interesting collective effects (for review articles see Refs. \cite{Duan:2010bg} and 
\cite{Duan:2009cd}). Such collective oscillations of neutrinos represent emergent nonlinear flavor evolution phenomena instigated by neutrino-neutrino interactions in astrophysical environments with sufficiently high neutrino densities. 

Sometimes  the term containing the angle between neutrino momenta is averaged over in an approximation known as the {\em single angle approximation}:
\begin{equation}
\label{satotal}
\hat{H}_{\mbox{\tiny SA}}   
= \left(
\sum_p\frac{\delta m^2}{2p}\mathbf{B}\cdot\mathbf{J}_p  - \sqrt{2} G_F 
N_e  J_p^0  \right) 
+ \frac{\sqrt{2}G_{F}}{V}   \left(\langle 1- 
\cos\vartheta_{\mathbf{p}\mathbf{q}} \rangle \right)  \sum_{\mathbf{p} \neq \mathbf{q}}     \mathbf{J}_{\mathbf{p}}\cdot\mathbf{J}_{\mathbf{q}}  .
\end{equation} 
A further approximation, which can be applied to either the multi-angle Hamiltonian of Eq. (\ref{total}) or the  single-angle Hamiltonian of Eq. (\ref{satotal}), is the mean field approximation. In this approximation the two-body term is replaced by a one-body term:
\begin{equation}
\mathbf{J}_{\mathbf{p}}\cdot\mathbf{J}_{\mathbf{q}} \rightarrow \langle \mathbf{J}_{\mathbf{p}} \rangle \cdot\mathbf{J}_{\mathbf{q}} 
+ \mathbf{J}_{\mathbf{p}}\cdot \langle  \mathbf{J}_{\mathbf{q}}  \rangle 
\end{equation}
where the averaging is done over an appropriately chosen state. Exactly how this state is chosen determines the physics that is emphasized. 
The evolution of the system under the many-body Hamiltonian of Eq. (\ref{total}) can be formulated as a coherent-state path integral,  
and a possible mean-field approximation represents the saddle-point solution of the path integral for this many-body system  
\cite{Balantekin:2006tg}. This is the most commonly used mean field approximation. One can also consider the contribution of neutrino-antineutrino pairing to the mean field \cite{Serreau:2014cfa}. Such a mean field would be proportional to the neutrino masses and could play a role in anisotropic environments \cite{Cirigliano:2014aoa}. 
 
\vskip 1cm 
\section{EXACT SOLUTIONS}

\subsection{Bethe Ansatz}

Exact solutions for the eigenstates of the Hamiltonian of Eq. (\ref{satotal}) are easier to calculate in the mass basis
\begin{equation}
\label{totalinmass}
\hat{H} 
= 
\sum_p \omega_p {\cal J}_p^0 + \mu  \sum_{\mathbf{p} \neq\mathbf{q}} \mathbf{\cal J}_{\mathbf{p}}\cdot\mathbf{\cal J}_{\mathbf{q}}  
\end{equation} 
where we introduced the neutrino flavor isospin operators in the mass basis:
\begin{eqnarray}
{\cal J}^+_{{\bf p}} &=& a_2^\dagger({\bf p}) a_1({\bf p}), \> \> \>
{\cal J}^{\>-}_{{\bf p}}=a_1^\dagger({\bf p}) a_2({\bf p}), \nonumber \\
{\cal J}^0_{{\bf p}} &=& \frac{1}{2}\left(a_2^\dagger({\bf p})a_2({\bf p})-a_1^\dagger({\bf p})a_1({\bf p}) 
\right), 
\label{su2mass}
\end{eqnarray}
with  
\begin{equation}
\omega_p = \frac{m_2^2 - m_1^2}{2p} = \frac{\delta m^2}{2p}
\end{equation}
and
\begin{equation}
\mu = \frac{\sqrt{2}G_{F}}{V} \left\langle \left(1- 
\cos\vartheta_{\mathbf{p}\mathbf{q}}\right) \right\rangle .
\end{equation}
In writing Eq. (\ref{totalinmass}) we ignored the electron background since we would like focus on regions where neutrino-neutrino interactions dominate. Note that the second term 
in this equation has the same form either in the mass or the flavor basis. 
Eigenvalues and eigenvectors of the Hamiltonian in Eq. (\ref{totalinmass}) can be calculated from the solutions of the Bethe ansatz 
equations \cite{Pehlivan:2011hp}:
\begin{equation}
\label{bethe}
-\frac{1}{\mu} - \sum_p \frac{j_p}{\omega_p - x_i} = \sum_{j \neq i}^N \frac{1}{x_i-x_j}
\end{equation}
In this equation $j_p$ is the SU(2) quantum number associated with the algebra describing neutrinos with momentum $p$. There are 
$2j_p+1$ such neutrinos. 
Once the solutions of the coupled equations in Eq. (\ref{bethe}) are identified, the eigenvectors are given by 
\begin{equation}
\label{eigenstate}
 \left( \sum_p \frac{a^{\dagger}_2(p) a_1(p)}{\omega_p - x_1} \right) \left( \sum_p \frac{a^{\dagger}_2(p) a_1(p)}{\omega_p - x_2} \right) \cdots \left( \sum_p \frac{a^{\dagger}_2(p) a_1(p)}{\omega_p - x_N} \right) \left( \prod_p a_1^{\dagger}(p) \right) | 0 \rangle 
\end{equation}
with $|0\rangle$ being the fermion vacuum, 
and the eigenvalues are given by 
\begin{equation}
E = 2 \sum_{p\neq q} j_p j_q - \sum_p \omega_p j_p- N\mu \sum_p j_p - \frac{\mu}{2} N(N-1) + \sum_{i=1}^N x_i .
\end{equation}

\vskip 0.5cm
\subsection{Spectral splits}

One interesting effect resulting from the collective neutrino oscillations is spectral swappings or splits, on the final neutrino energy spectra: at a particular energy these spectra are almost completely divided into parts of different flavors \cite{Raffelt:2007cb,Duan:2008za}. In the single-angle limit of the full many-body Hamiltonian it was shown that the spectral split of a neutrino ensemble, which initially consists of single flavor neutrinos, is analogous to the crossover from the BCS to the Bose-Einstein condensate limits \cite{Pehlivan:2016lxx}. 
When one uses the mean field approximation for the two-flavor case, total neutrino number is no longer conserved and needs to be enforced using a Lagrange multiplier. This Lagrange multiplier can be interpreted as the critical energy where the spectral swap/split takes place  \cite{Pehlivan:2011hp}. These swaps may be independent of the mean field approximation:  
In Ref. \cite{Pehlivan:2016voj} the adiabatic evolution of a full many-body state of 250 electron neutrinos with inverted hierarchy distributed in a thermal energy spectrum was followed as the value of $\mu$ decreased from a very high value down to zero. It was found that the resulting split energy is the same as that was obtained in the mean-field approximation. 

To illustrate the spectral split behavior in the full-many body case, we consider a simple toy model with two neutrinos ($j_1 = j_2 = 1/2$) with momenta yielding values $\omega_1$ and 
$\omega_2$. For this simple case the Bethe ansatz equations, Eq. (\ref{bethe}) can be solved providing two different solutions for $N=1$:
\begin{equation}
\label{N1sol}
x_{\pm} = \frac {\omega_1 + \omega_2}{2} + \frac{1}{2} \mu \pm \frac{1}{2} \sqrt{(\omega_1-\omega_2)^2 + \mu^2} ,
\end{equation}
and a pair of solutions for $N=2$:
\begin{equation}
\label{N2sol}
x_{1,2} = \frac {\omega_1 + \omega_2}{2} + \frac{1}{2} \mu \pm \frac{1}{2} \sqrt{(\omega_1-\omega_2)^2 - \mu^2} .
\end{equation}
Note that the latter solutions become complex as $\mu$ gets very large. Let us assume that $\omega_1 > \omega_2>0$ and consider $N=1$ case. 
\begin{figure}[t]
\label{f:1}
\caption{Solutions of the Bethe ansatz equations given in Eq. (\ref{N1sol}). Upper two lines are $x_+$ for $\omega_2/\omega_1 = 0.7$ (upper thick solid line) and  $\omega_2/\omega_1 = 0.3$ (upper thin solid line) wheras lower two lines are $x_-$ for $\omega_2/\omega_1 = 0.7$ (lower thick dashed line) and  $\omega_2/\omega_1 = 0.3$ (lower thin dashed line).}
\includegraphics[height=.3\textheight]{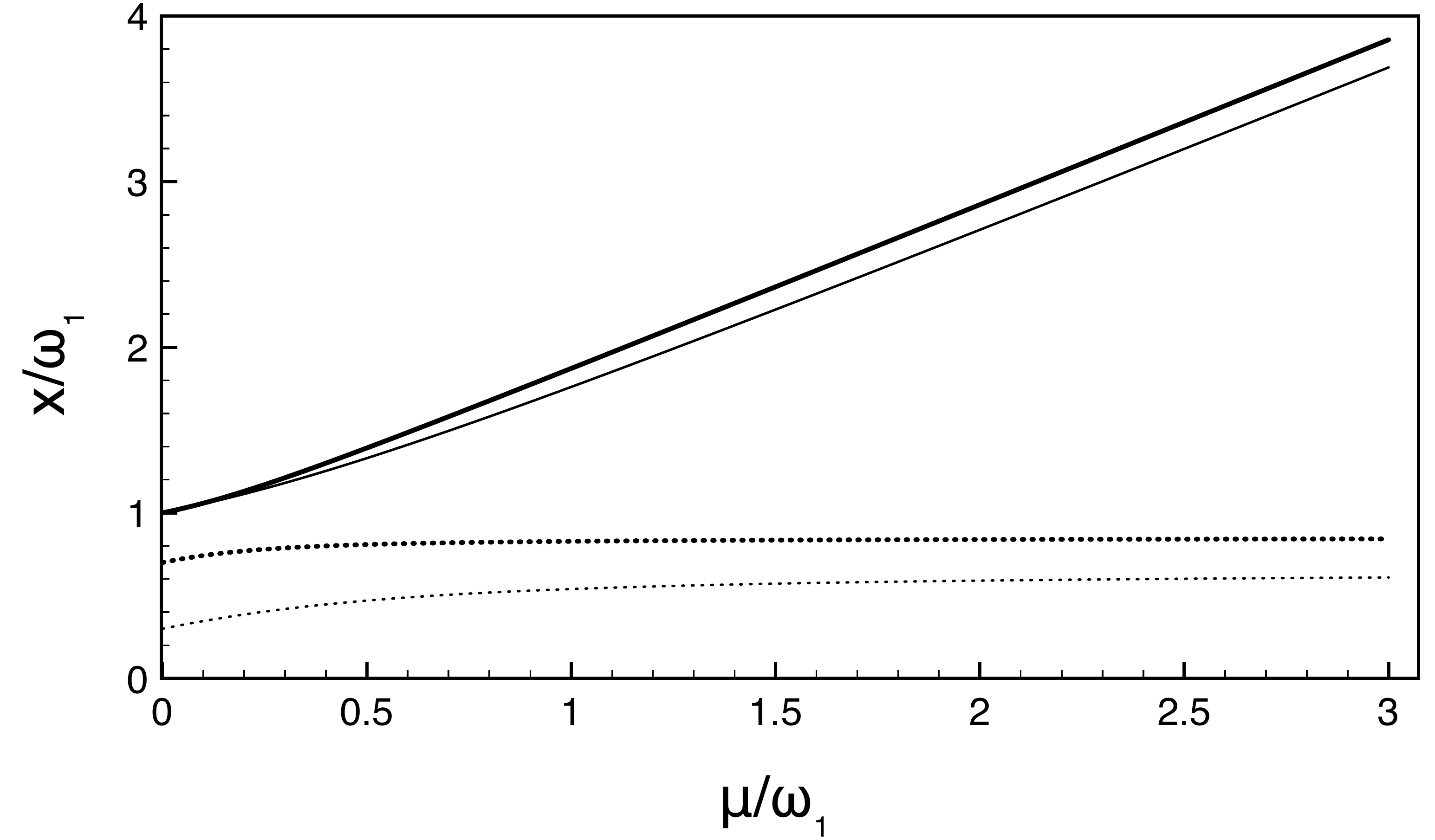}
\end{figure}
In Figure 1 we show solutions given in Eq. (\ref{N1sol}): upper two lines denote $x_+$ for $\omega_2/\omega_1 = 0.7$ (upper thick solid line) and  $\omega_2/\omega_1 = 0.3$ (upper thin solid line) wheras lower two lines denote $x_-$ for $\omega_2/\omega_1 = 0.7$ (lower thick dashed line) and  $\omega_2/\omega_1 = 0.3$ (lower thin dashed line). This figure illustrates what seems to be a generic property of the solutions of the Bethe ansatz equations: there is one solution that increases with increasing $\mu$ eventually becoming infinite  whereas the other solution remains finite. The solution which grows with $\mu$ is the solution which starts as $\omega_1$ at $\mu=0$ (other solution 
starts as $\omega_2$ at $\mu=0$). 
Note that $\omega_1$ is the larger of the two 
$\omega$ values and corresponds to the smaller of the two values of neutrino momenta. For example, in the case of the adiabatic expansion of the neutrino gas formed in a core-collapse supernova $\mu \rightarrow \infty$ represents the neutrinosphere region and $\mu \sim 0$ represents the outer shells. Hence the behavior of the solutions described above gives us a hint on how neutrinos may exchange energy as they travel away from the neutrinosphere. Here to be specific let us consider the adiabatic eigenstates. 
The normalized 
eigenstates are given by 
\begin{equation}
| \xi_+ \rangle = \frac{1}{\cal N} \left(  \frac{2 a_2^\dagger({\bf p_1}) a_1({\bf p_1})}{\eta - \mu- \sqrt{\eta^2 + \mu^2}} - \frac{2 a_2^\dagger({\bf p_2}) a_1({\bf p_2})}{\eta + \mu +\sqrt{\eta^2 + \mu^2}} \right) a_1^{\dagger} (p_1) a_1^{\dagger}(p_2) | 0 \rangle 
\end{equation}
with $\eta = \omega_1 - \omega_2$ and 
\begin{equation}
{\cal N}^2 = 4 \left( \frac{1}{(\eta - \mu- \sqrt{\eta^2 + \mu^2})^2} + \frac{1}{\eta + \mu +\sqrt{\eta^2 + \mu^2}} \right). 
\end{equation}
One can write a similar expression for $|\xi_-\rangle$. One can easily show that as $\mu \rightarrow 0$ 
\begin{eqnarray}
\lim_{\mu \rightarrow 0} |\xi_+ \rangle &=&  {\cal J}^+ (p_1) |0 \rangle,  \nonumber \\
\lim_{\mu \rightarrow 0} |\xi_- \rangle &=&  {\cal J}^+(p_2) |0 \rangle 
\end{eqnarray}
which are the eigenstates of the first term in Hamiltonian of Eq. (\ref{totalinmass})  
as expected. Similarly
\begin{eqnarray}
\lim_{\mu \rightarrow \infty} |\xi_+ \rangle &=& \frac{1}{\sqrt{2}} \left( {\cal J}^+ (p_1) + {\cal J}^+ (p_2)\right) |0 \rangle,  \nonumber \\
\lim_{\mu \rightarrow \infty} |\xi_- \rangle &=& \frac{1}{\sqrt{2}} \left( {\cal J}^+ (p_1) -  {\cal J}^+ (p_2) \right) |0 \rangle. 
\end{eqnarray}
As an example let us consider the state
\begin{equation}
| \psi (\mu) \rangle = \frac{1}{\sqrt{2}} \left( | \xi_+ \rangle + | \xi_- \rangle \right) .
\end{equation}
From the expressions above we see that near the neutrinosphere such a state would be 
\begin{equation}
| \psi (\mu \rightarrow \infty) \rangle = a_2^{\dagger} (p_1) a_1^{\dagger} (p_2) | 0 \rangle, 
\end{equation}
i.e. the neutrino in the mass eigenstate $2$ has momentum $p_1$ and the neutrino in the mass eigenstate $1$ has momentum $p_2$.  Far away from the proto-neutron star this state would take the form 
\begin{equation}
| \psi (\mu \rightarrow 0) \rangle = \frac{1}{\sqrt{2}} \left( a_2^{\dagger} (p_1) a_1^{\dagger} (p_2) \> | 0 \rangle -   a_2^{\dagger} (p_2) a_1^{\dagger} (p_1) \> | 0 \rangle \right) ,
\end{equation}
i.e. either mass eigenstate has equal probability to carry momenta $p_1$ and $p_2$. One should emphasize that, since the above arguments 
assume $\omega_1 > \omega_2$, the evolution will be opposite in the inverted hierarchy case (when the signs of $\omega$s change) than in the normal hierarchy case. 

\vskip 0.5cm
\subsection{Conserved quantities}

It can be shown that there are additional conserved quantities commuting with the Hamiltonian in Eq. (\ref{totalinmass}). They can be written in terms of quantity \cite{Pehlivan:2011hp}
\begin{equation}
\label{hpinv}
\hat{h}_p  =  {\cal J}_p^0 + \mu  \sum_{q, \mathbf{q} \neq\mathbf{p}} 
\frac{\mathbf{\cal J}_{\mathbf{p}}\cdot\mathbf{\cal J}_{\mathbf{q}}}{\omega_p - \omega_q}  .
\end{equation} 
In fact the Hamiltonian can be written in terms  of these quantities:
\begin{equation}
\hat{H} = \sum_p \omega_p \hat{h}_p. 
\end{equation}
The eigenvalues of $\hat{h}_p$ when acted on the state in Eq. (\ref{eigenstate}) are given by 
\begin{equation}
\epsilon_p = \mu \sum_{q, q\neq p} \frac{j_pj_q}{\omega_p - \omega_q} - j_p -\mu j_p \sum_{i=1}^N \frac{1}{\omega_p - x_i} .
\end{equation}

\vskip 0.5cm
\subsection{Three flavors and CP Violation}

In the previous sections we ignored the third flavor state as well as antineutrinos. 
A third flavor can be incorporated by introducing SU(3) as the neutrino flavor isospin algebras for each momenta and antineutrinos require a second set of the SU(3) algebras. 
One can investigate the symmetries of the problem in the full three flavor mixing scheme and in the exact many-body formulation, and in addition include the effects of CP violation and neutrino magnetic moments \cite{Pehlivan:2014zua}. One finds that, similar to what was discussed above, several dynamical symmetries exist for three flavors in the single-angle approximation if the net electron background in the environment and the effects of the neutrino magnetic moment are negligible. These dynamical symmetries are present even in the presence of CP-violating phases. One can explicitly write down the constants of motion through which these dynamical symmetries manifest themselves in terms of the generators of the SU(3) flavor transformations. In this case the effects due to the CP-violating Dirac phase factor out of the many-body evolution operator and evolve independently of nonlinear flavor transformations if neutrino electromagnetic interactions with external magnetic fields are ignored \cite{Pehlivan:2014zua}. 

\vskip 1cm
\section{COLLECTIVE OSCILLATIONS FOR THE $\nu$p PROCESS}

An example of the application of multi-angle three-flavor calculations was recently given in Ref. \cite{Sasaki:2017jry} where the effects of collective neutrino oscillations on $\nu$p process nucleosynthesis in proton-rich neutrino driven winds were studied by coupling neutrino transport calculations with nucleosynthesis network calculations.  Collective neutrino oscillations transform the spectra of all neutrino species, but of particular importance is the modification of the electron neutrino and electron antineutrino energy distributions: the capture of $\nu_e$ and $\bar{\nu}_e$ on free nucleons determine the neutron-to-proton ratio, hence the yields of nucleosynthesis. 

In simple spectral split scenarios motivated by the single-angle collective oscillations abundances of p-nuclei are enhanced when the outflows are proton rich \cite{MartinezPinedo:2011br}. However the single-angle approximation, since it ignores angular correlations, changes the onset of the collective oscillations. For example, nucleosynthesis yields are drastically different between single-angle and multi-angle 
calculations of the supernova r-process \cite{Duan:2010af}. A complete spectral swap as obtained in calculations with the single-angle approximation do not seem to emerge from the multi-angle calculations: in both hierarchies the onset of collective oscillations are delayed as compared with that in the single angle approximation. 

In Ref. \cite{Sasaki:2017jry} calculations were carried out for two  proton-rich neutrino-driven winds at $t=0.6$ s and $t=1.1$ s after the core bounce in a one-dimensional explosion simulation model. For  the early wind trajectory collective neutrino oscillations in the inverted mass hierarchy increase the energetic $\nu_e$ flux in the region where the $\nu$p nucleosynthesis takes place. However, in the later wind 
trajectory  oscillations increase energetic $\bar{\nu}_e$s leading to an enhancement of the $\nu$p process, about up to 20 times larger than that was obtained in Ref. \cite{MartinezPinedo:2011br}. The fact that the enhancement is dominated by the later wind suggests that model dependence of these results may not be too strong.

Of course uncertainties of the initial neutrino fluxes as well as hydrodynamic quantities such as the wind velocity could alter these conclusions. 
If the luminosities and the energies of the neutrino fluxes for different flavors are similar, then the effects of the neutrino oscillations would be minimized. However, it is clear that collective neutrino oscillations {\it can} significantly influence the $\nu$p process yields and further systematic studies of neutrino physics and hydrodynamics input would be very useful.

\section{ACKNOWLEDGMENTS}
This work was supported in part by the US National Science 
Foundation Grant No. PHY-1514695 and and 
in part by the University of Wisconsin Research Committee with funds 
granted by the Wisconsin Alumni Research Foundation. 
I would like to thank National Astronomical Observatory of Japan for its hospitality and  acknowledge my collaborators T. Kajino, Y. Pehlivan, T. Hayakawa, H. Sasaki, T. Takiwaki, and T. Yoshida 
who contributed to much of the work reported here. 


\nocite{*}
\bibliographystyle{aipnum-cp}%
\bibliography{sample}%

\begin{thebibliography}{99}

\bibitem{Balantekin:2006tg} 
  A.~B.~Balantekin and Y.~Pehlivan,
  J.\ Phys.\ G {\bf 34}, 47 (2007)
  [astro-ph/0607527].
  
\bibitem{Duan:2010bg} 
  H.~Duan, G.~M.~Fuller and Y.~Z.~Qian,
  Ann.\ Rev.\ Nucl.\ Part.\ Sci.\  {\bf 60}, 569 (2010)
  [arXiv:1001.2799 [hep-ph]].
  
\bibitem{Duan:2009cd} 
  H.~Duan and J.~P.~Kneller,
  J.\ Phys.\ G {\bf 36}, 113201 (2009)
  doi:10.1088/0954-3899/36/11/113201
  [arXiv:0904.0974 [astro-ph.HE]].
  
\bibitem{Serreau:2014cfa} 
  J.~Serreau and C.~Volpe,
  Phys.\ Rev.\ D {\bf 90}, 125040 (2014)
  [arXiv:1409.3591 [hep-ph]].
  
\bibitem{Cirigliano:2014aoa} 
  V.~Cirigliano, G.~M.~Fuller and A.~Vlasenko,
  Phys.\ Lett.\ B {\bf 747}, 27 (2015)
  [arXiv:1406.5558 [hep-ph]].
  
\bibitem{Pehlivan:2011hp} 
  Y.~Pehlivan, A.~B.~Balantekin, T.~Kajino and T.~Yoshida,
  Phys.\ Rev.\ D {\bf 84}, 065008 (2011)
  [arXiv:1105.1182 [astro-ph.CO]].
  
\bibitem{Raffelt:2007cb} 
  G.~G.~Raffelt and A.~Y.~Smirnov, 
  Phys.\ Rev.\  D {\bf 76}, 081301 (2007) 
  [Erratum-ibid.\  D {\bf 77}, 029903 (2008)] 
  [arXiv:0705.1830 [hep-ph]]; 
  Phys.\ Rev.\  D {\bf 76}, 125008 (2007) 
  [arXiv:0709.4641 [hep-ph]]. 
 
\bibitem{Duan:2008za}
H. Duan, G. M. Fuller, Y.-Z. Qian,
Phys. Rev. D {\bf 77}, 085016 (2008) 
[arXiv:0801.1363 [hep-ph]].
  
\bibitem{Pehlivan:2016lxx} 
  Y.~Pehlivan, A.~L.~Subasi, N.~Ghazanfari, S.~Birol and H.~Yuksel,
  Phys.\ Rev.\ D {\bf 95}, no. 6, 063022 (2017)
  doi:10.1103/PhysRevD.95.063022
  [arXiv:1603.06360 [astro-ph.HE]].
  
\bibitem{Pehlivan:2016voj} 
  Y.~Pehlivan, A.~L.~Subasi, S.~Birol, N.~Ghazanfari, H.~Yuksel, A.~B.~Balantekin and T.~Kajino,
  AIP Conf.\ Proc.\  {\bf 1743}, 040007 (2016).
  doi:10.1063/1.4953299
 
\bibitem{Pehlivan:2014zua} 
  Y.~Pehlivan, A.~B.~Balantekin and T.~Kajino,
  Phys.\ Rev.\ D {\bf 90}, no. 6, 065011 (2014)
  doi:10.1103/PhysRevD.90.065011
  [arXiv:1406.5489 [hep-ph]].
 
\bibitem{Sasaki:2017jry} 
  H.~Sasaki, T.~Kajino, T.~Takiwaki, T.~Hayakawa, A.~B.~Balantekin and Y.~Pehlivan,
  Phys.\ Rev.\ D {\bf 96}, no. 4, 043013 (2017)
  doi:10.1103/PhysRevD.96.043013
  [arXiv:1707.09111 [astro-ph.HE]].
  
\bibitem{MartinezPinedo:2011br} 
  G.~Martinez-Pinedo, B.~Ziebarth, T.~Fischer and K.~Langanke,
  Eur.\ Phys.\ J.\ A {\bf 47}, 98 (2011)
  doi:10.1140/epja/i2011-11098-y
  [arXiv:1105.5304 [astro-ph.SR]].
  
\bibitem{Duan:2010af} 
  H.~Duan, A.~Friedland, G.~C.~McLaughlin and R.~Surman,
  J.\ Phys.\ G {\bf 38}, 035201 (2011)
  doi:10.1088/0954-3899/38/3/035201
  [arXiv:1012.0532 [astro-ph.SR]].
    
   
\end{thebibliography}

\end{document}